# Optical Control of Adaptive Nanoscale Domain Networks

Marc Zajac[1], Tao Zhou[2], Tiannan Yang[3,11], Sujit Das[4,12], Yue Cao[5], Burak Guzelturk[1], Vladimir Stoica[3], Mathew Cherukara[1], John W. Freeland[1], Venkatraman Gopalan[3], Ramamoorthy Ramesh[4,7,8,9], Lane W. Martin[4,7,8,9,10], Long-Qing Chen[3], Martin Holt[2], Stephan Hruszkewycz[5*], Haidan Wen[1,5*]

[1] Advanced Photon Source, Argonne National Laboratory, Lemont, IL 60439
[2] Center for Nanoscale Materials, Argonne National Laboratory, Lemont, IL 60439
[3] Department of Materials Science and Engineering, Pennsylvania State University, University Park, PA 16802
[4] Materials Sciences Division, Lawrence Berkeley National Laboratory, Berkeley, CA 94720
[5] Materials Science Division, Argonne National Laboratory, Lemont, IL 60439
[7] Department of Materials Science and NanoEngineering, Rice University, Houston, TX 77005
[8] Department of Physics and Astronomy, Rice University, Houston, TX 77005
[9] Rice Advanced Materials Institute, Rice University, Houston, TX 77005
[10] Department of Chemistry, Rice University, Houston, TX 77005
[11] Present address: Interdisciplinary Research Center, School of Mechanical Engineering, Shanghai Jiao Tong University, Shanghai, 200240, China
[12] Present address: Materials Research Centre, Indian Institute of Science, Bangalore, 560012, Karnataka, India

[*]Email: shrus@anl.gov; wen@anl.gov

**Abstract**

Adaptive networks can sense and adjust to dynamic environments to optimize their performance. Understanding their nanoscale responses to external stimuli is essential for applications in nanodevices and neuromorphic computing. However, it is challenging to image such responses on the nanoscale with crystallographic sensitivity. Here, the evolution of nanodomain networks in $(PbTiO_3)_n/(SrTiO_3)_n$ superlattices was directly visualized in real space as the system adapts to ultrafast repetitive optical excitations that emulate controlled neural inputs. The adaptive response allows the system to explore a wealth of metastable states that were previously inaccessible. Their reconfiguration and competition were quantitatively measured by scanning x-ray nanodiffraction as a function of the number of applied pulses, in which crystallographic characteristics were quantitatively assessed by assorted diffraction patterns using unsupervised machine-learning methods. The corresponding domain boundaries and their connectivity were drastically altered by light, holding promise for light-programmable nanocircuits in analogy to neuroplasticity. Phase-field simulations elucidate that the reconfiguration of the domain networks is a result of the interplay between photocarriers and transient lattice temperature. The demonstrated optical control scheme and the uncovered nanoscopic insights open opportunities for remote control of adaptive nanoscale domain networks.

## 1. Introduction

The ability to sense and adapt to dynamic external stimuli is a unique signature of smart systems. Biological neural networks possess sufficient complexity for connectivity, flexibility to adapt, and a large enough configurational phase space to respond to these stimuli. Neuromorphic computing systems attempt to emulate such biological systems through solid-state analogs. Although complementary metal–oxide–semiconductor-based artificial neurons are commercially available, non-von Neumann approaches will require new material systems that possess not only memory retention but also flexibility in adaptive reconfiguration as in neural networks for next-generation computation hardware[1–6].

Ferroelectric materials exhibit non-volatile polarization states that can be switched electrically. Their dynamic response to electric fields can mimic certain aspects of biological neurons such as synaptic plasticity, which is crucial for learning and memory in biological neural networks[7,8]. Recently, the properties of ferroelectrics have been enriched by the discovery of emergent polar textures in $(PbTiO_3)_n/(SrTiO_3)_n$ (PTO/STO) superlattices (SLs), providing larger phase space for reconfigurability. Alternating epitaxial layers of PTO and STO each with a thickness of $n$ unit cells on a single-crystal substrate can host nanoscale arrangements of polar vortices[9,10] or skyrmions[11] that give rise to unique physical phenomena including chirality and circular dichroism[12], sub-terahertz collective modes[13], and negative permittivity[14]. By tuning the lattice strain and applying electrical fields, a larger diversity of polar nanostructures has been observed[10,15–18]. In addition, light-matter interactions offer avenues not only for switching polarizations in ferroelectric materials[19–22] but for creating statesthat are difficult to achieve via conventional, thermodynamic-equilibrium pathways[23], and can exhibit stochastic responses[24].

Understanding light-induced states, either transient[25–32] or metastable[23,33–35], however, is complicated by nanoscale structures such as polar nanoregions[33], topological defects[25], and phase separations[30–32], which are challenging to visualize and characterize under optical excitation. In this respect, PTO/STO SLs are a model system to study the emergence and the interplay of light-induced adaptive responses of nanostructures and their domain networks. First, the system is poised by precision synthesis at the verge of phase transitions and is thus strongly responsive to external stimuli. Upon subtle but cumulative excitations that emulate trains of neural inputs, the adaptability of the nanoscale domain network can be systematically studied. The system reconfigures its domain network including changing the domain size and connectivity for stability according to the level of external stimuli. Second, a wealth of polar nanostructures with distinct polarization arrangements, hereafter referred to as a "phase", can be spatially resolved and quantitatively characterized by x-ray diffraction imaging. These metastable phases and their nanoscale networks that can be spatially resolved provide a platform to quantitatively study adaptive responses analogous to neuron networks.

Here, we demonstrated optical control of domain networks and their phase configurations in PTO/STO SLs and systematically imaged nanoscale responses as a function of gradually building above-band-gap optical excitations by x-ray nanodiffraction imaging (Figure 1a and Methods). Metastable polar phases were created, erased, or transformed by trains of ultrafast light pulses (*i.e.*, using a repetitive ultrafast excitation scheme). The regions and boundaries of these phases were found to evolve and rearrange with length scales from 10 nm to 10 µm. Our results show that a repetitive excitation protocol can explore a wider range of the "light-induced" phase diagram (Figure 1b) and access a wealth of new polarization configurations (insets, Figure 1b) that are detailed later. The use of above-band-gap femtosecond pulse trains takes advantage of cumulative excitations to create significantly higher trapped carrier concentrations while keeping the peak-lattice temperature lower than what is typical of intense single-shot excitation. Unlike the intense single-shot laser pulse that transforms it into a "three-dimensional supercrystal"[23] (red arrow, Figure 1b), this strategy enables the system to adapt to controlled stimulation, unlocking pathways to new nanoscale phases (blue arrows, Figure 1b). In addition, advanced nanodiffraction imaging and the corroborative phase-field simulations employed in our work reveal the nanoscale response of domains and their boundaries as well as mechanistic insights into optical control of nanodomain networks.

## 2. Evolution of Nanoscale Domains and Boundaries in Real Space

The *in-situ* imaging experiments with optical excitation were performed at the hard x-ray nanoprobe beamline of the Center for Nanoscale Materials (Methods). $(PTO)_{16}/(STO)_{16}$ SLs grown on $DyScO_3$ substrates (Methods) were exposed to trains of femtosecond pulses with a wavelength of 343 nm at a repetition rate of 67 kHz. This wavelength was chosen because it is above the band gaps of PTO and STO. The exposure time (*i.e.*, the duration of the pulse train) is controlled by a mechanical shutter from 1-1000 s (Figure 1a). The absorbed photon density per shot is about ten times smaller than the threshold required for creating the previously observed light-induced supercrystal phase[23]. In pristine PTO/STO SLs[10], streaks of the vortex phase alternate with streaks of the in-plane polarized ($a_1/a_2$) phases, forming in-plane orders with a periodicity of ≈400 nm along $[010]_{PTO} \parallel [\bar{1}10]_{DSO}$, defined as the direction of the *y* axis throughout. This alternating phase pattern was resolved in real space by measuring the intensity of the Bragg peaks from the vortex phase using a scanning nanofocused x-ray diffraction probe (Figure 1a, c). The in-plane nanoscale ordering of polarization domains within these phases can be measured by analyzing diffuse satellite diffraction peaks (Figure 1d) which originate from vortex tubes with an ≈11 nm period[9,23].

After each exposure, the nanofocused x-ray beam was scanned across the sample at an incident angle of $\theta$ =18.3°. This incident angle satisfied the Bragg condition of the superlattice ($m$ = -1) peak of the vortex phase along the 00$L$ specular crystal truncation rod (CTR), where $m$ represents the order number of the superlattice reflection and $L$ is the Miller index. The Bragg peaks appeared as annulus-shaped patterns, as is typical with zone-plate-focused x-ray nanodiffraction (Methods). The out-of-plane lattice constants of the vortex and $a_1/a_2$ phases are slightly different (Table 1, Supporting Information), which leads to the separation of their diffraction peaks along the $Q_z$' axis on the detector. The diffraction intensity of the $m$ = -1 SL vortex peak integrated within the white box (Figure 1d) is plotted as a function of real-space positions (Figure 1c). After an optical exposure of 10 s (*i.e.*, 6.7×10$^5$ pulses), patches containing horizontally aligned streaks (period of ≈280 nm) emerged from the vertically aligned streaks (period of ≈400 nm). Such a change in the mesoscale order indicates that a complex evolution of polar phases can be controlled by tuning the optical dosage.

The localized nature of the focused beam (35 nm in diameter) allows diffraction features that cannot be separated in ensemble-averaged measurements to be resolved spatially. Figure 2a shows the evolution of the real-space distribution of the $m$ = -1 SL peak intensity as a function of optical exposure time. Horizontal streaks initially arose after the 5 s exposure and grew after the subsequent 10 s exposure. Further optical doses of 50-100 s converted more regions into similar horizontal streaks. In the maps after 500 s and 1000 s exposures, the optically induced changes were saturated so that further exposure did not lead to new polar phases. In this regime, the streak textures were mostly replaced by micrometer-sized areas with little intensity variation. In the images after a 1000 s exposure, low-intensity, diagonal features and dark spots in the map were visible and can be traced back to the pristine state. These dose-invariant features were used as fiducials to determine a constant field of view to account for position drift (Figure S1, Supporting Information).

Deeper insights into the nanoscale ordering of the emerging polar phases were gained by analyzing the scattering features in a much wider reciprocal space window (Figure 2b) rather than the $m$ = -1 SL peak alone. Unsupervised machine-learning techniques[36,37] were used to classify the diffraction patterns into five categories (Methods). The five classes of diffraction patterns correspond to polar phases with distinct nanoscale ordering that we refer to as 1) vortex, 2) $a_1/a_2$, 3) vortex-like-xy (VL-xy), 4) super-stripe-x (SS-x), and 5) super-stripe-y (SS-y). Schematics of the proposed real-space polar structures based on phase-field simulations are displayed as insets of Figure 1b. The first two phases are known to be present in the as-grown sample[10], while the VL-xy and SS phases have not been previously reported. The VL-xy phase contains vortex-like tubes that are aligned along either the $x$ or $y$ axis. This is different from as-grown vortex tubes that are only ordered along the $y$ axis, as well as $a_1/a_2$ phase that display in-plane polarization and are ordered along 45° direction with respect to the $x$ and $y$ axes (Figure 1c). SS-x and SS-y phases occupy micron-sized patches with a stripe-ordered periodicity that is double the vortex phase periodicity and preferentially aligned along either the $x$ or $y$ axis, unlike the previously reported 3D supercrystal phase that orders along all three dimensions[23]. These proposed models for the nanoscale polarization domain arrangements are verified later by three-dimensional reciprocal space maps.

We first discuss the evolution of real-space coverage of the five phases (Figure 2c), which share the same color coding as in Figure 2b. In the pristine state, both vortex and $a_1/a_2$ phases were dominant, because the pure $a_1/a_2$ phase or the vortex domains are not stable at room temperature due to the large tensile or compressive epitaxial strain associated with each phase along the *y* axis ( ∥ $[\bar{1}10]_{DSO}$), respectively (Table 1, Supporting Information). Spatial arrangements in which vortex and $a_1/a_2$ phases alternate along the *y* axis, however, can reduce the average strain-gradient energy over hundreds of nanometers to stabilize both phases[9,23]. The phase composition and boundaries were found to evolve in the following three optical-dose regimes.

First, in the *low-dose* regime (1 s exposure), some of the initial $a_1/a_2$ phase was converted into the vortex phase, indicating that a low optical dosage can generate appropriate carrier concentrations to promote the formation of vortex domains. The increase of the vortex-phase fraction contrasts with the electric-field-driven transition from vortex to $a_1/a_2$ phases[10] and single-shot optical excitation[23], highlighting the unique capability of the cumulative optical excitation scheme. These changes, as suggested by the phase-field simulation[16] but not yet observed previously, mainly occur in regions with high electrostatic energy such as the charged PTO/STO interfaces and domain walls to compensate for the net bound polarization charges of the polar structure.

Second, as the optical dose continued to increase into the *intermediate-dose* regime (5-10 s exposure), the VL-xy phase (light green, Figure 2c) that appeared initially after the 1 s dose grew further in regions containing both $a_1/a_2$ and vortex phases. As will be shown later in Figure 3, the VL-xy phase features in-plane modulation of vortex-like tubes along both the *x* and *y* axes, rather than only along the *x* axis. Once nucleated, the VL-xy phase grew mostly isotropically, subsuming both the $a_1/a_2$ and the vortex phases. This observation supports the notion that the strain gradient energies of VL-xy are similar in both directions. The VL-xy phase is likely composed of VL-x and VL-y phases that coexist within the probe volume (Figure S2, Supporting Information), and thus they cannot be resolved individually in our measurements.

Finally, the *high-dose* regime (50-1000s exposure) was marked by the appearance and redistribution of SS-x and SS-y domains, and the complete disappearance of the $a_1/a_2$ phase. After 100 s of exposure, the vortex phase that is characterized by large in-plane lattice anisotropy was replaced by VL-xy and SS phases. The phases present in this regime featured reduced out-of-plane strain gradients across phase boundaries, as evidenced by the similar *c* lattice constant among these domains (color bar of Figure 2c). The reduction of ferroelastic energy within the VL-xy, SS-x, and SS-y phases allows them to grow into much larger patches. No significant changes were observed upon further increasing the exposure time from 500 to 1000s. In this regime, the VL-xy phase occupied a majority of the field of view because its biaxial ordering minimized the epitaxial strain. The SS-x and SS-y phases occupy micron-sized patches, which can be distinguished by the characteristic nanodiffraction patterns of the *x*- and *y*-aligned stripe variants of the SS phase (Figure S3, Supporting Information). By tracking the domain evolution in real space, non-monotonic changes of the phase fractions were observed (black circled areas, Figure 2c). Since the probed region is much smaller than the pump beam size (200 µm), the pump intensity is homogeneous. The observed interconversion of phases indicates a stochastic process, as the VL-xy, SS-x, and SS-y species are close in the free energies (Table 2, Supporting Information) and transmute.

After repeating the experiment with a reduced optical fluence of 1.15 mJ cm$^{-2}$ (Figure S4, Supporting Information), a longer exposure time was needed to drive the phase transformation. The VL-xy phase was observed after 50 s of exposure, about ten times longer than that was needed at a fluence of 2.3 mJ cm$^{-2}$ (Figure 2c), although the fluence per pulse was only reduced by a factor of two. This indicates a nonlinear polarization switching regime. Using the nucleation-limited switching model in ferroelectrics[38], one can estimate the wait time $\tau$ for nucleation, which scales with the applied field $E$ as $\tau \propto \exp\left(\frac{1}{E}\right)^{1.5}$. Since the photocarrier concentration is proportional to the internal depolarizing field[23], the factor-of-two reduction in carrier concentration results in seventeen times longer wait time $\tau$ for nucleation, which is in approximate agreement with our experimental observation. Besides the demonstrated controllability of phase transition by tuning the fluence, other excitation parameters such as pulse interval, duration and wavelength can potentially offer more controllability in the multi-shot excitation scheme.

To obtain insights into the photo-induced domain reconfiguration, we analyzed the evolution of the phase boundaries. Overall, the spatial density of phase boundaries significantly reduced as the optical dose increased (Figure S5, Supporting Information). We demonstrated this trend by visualizing the changes in Figure 2d, where changes in the boundaries between exposure times were plotted. This was based on Figure S5 in Supporting Information, which shows the boundaries after consecutive exposures. These differential boundary maps highlight boundaries that were either erased, generated, or preserved. Most of the original boundaries were erased by the 50 s exposure (high coverage of dark blue contours) as the system favors creating large patches of SS-x or -y domains. Some parallel pairs of erased/created domain boundaries indicate that photoexcitation also promotes domain-boundary migration due to modified stress at the domain boundaries[39]. We also mapped the pinned-phase boundaries (left panel, Figure 2d) and found that only 1% of the original boundary locations persisted through the entire optical dose series. The ability to drastically reconfigure the phase boundaries is in contrast to the recovery of domain boundaries in $BiFeO_3$ upon optical excitation[40], suggesting possible applications of PTO/STO SLs in programable nanocircuits using light. The boundaries in the 1000 s image are mostly aligned along the 45° direction with respect to the crystalline axis, as needed for joining the in-plane stripe domains of the SS-x and -y phases.

The light-induced phases discussed above were created by ultrafast laser excitation at room temperature. Their structural stability lasts for months, as evident by the same diffraction patterns measured in a follow-up nano-XRD synchrotron experiment scheduled four months later. These light-induced phases are not permanent but metastable since the sample can recover to the pristine state by performing a thermal cycle from room temperature to 500 K.

## 3. Characterization of Nanoscale Domains in Reciprocal Space

Having discussed the evolution of polar phases at the mesoscale, we turn our attention to discussing the diffraction features that support the proposed models of the nanoscale polarization domain ordering of each phase. Since the nanodiffraction maps discussed above were taken at a fixed angle, only a two-dimensional slice through the three-dimensional reciprocal space was measured (Figure S6a, Supporting Information). To provide three-dimensional reciprocal space maps (RSM), we performed rocking scans with the nanofocused x-ray beam (Method). The RSM of the pristine state was consistent with the presence of vortex and $a_1/a_2$ phases, as has been previously reported[10], featuring the first-order satellite peaks offset from the central CTR along the $Q_x$ axis. There were only weak satellite peaks along the $Q_y$ direction, indicating that the vortex tubes that extend along the *y* axis in the pristine state are predominantly ordered along the *x* axis, which is oriented along [001]$_{DSO}$.

The RSM of the VL-xy phase differed from the pristine state. Along the out-of-plane $Q_z$ direction, the satellite peaks were less diffuse than those in the pristine state (the yellow box in the middle panel of Figure 3a), indicating that the VL-xy polarization domains had longer correlation lengths along the out-of-plane direction. In the $Q_x$-$Q_y$ cut of the RSM, satellite peaks appeared in both the $Q_x$ and $Q_y$ directions with similar reciprocal-space spacing, indicating similar in-plane ordering of polar domains along both directions. The larger in-plane periodicity of this peak corresponds to ≈12 nm and is consistent with the reduction of the out-of-plane lattice constant of the VL-xy state (3.931 Å), measured by the Bragg peak position (yellow arrows, Figure 3a).

The RSM of the SS phase included diffraction from both SS-x and -y domains and featured new diffraction peaks corresponding to nanoscale-ordered stripes with periodicities that are close to double that of the VL-xy phase. The first ($k = 1$) and second ($k = 2$) order peaks appear at different $Q_z$, a feature also seen previously in the three-dimensional supercrystal phase. The corresponding in-plane periodicity of the SS polar domains are 29 nm and 25 nm for the SS-x and SS-y variants, respectively, similar to the in-plane periodicity of 30 nm and 25 nm of the three-dimensional supercrystal phase[23]. The SS-x and -y phase domains, however, do not interlace to form the checkerboard-type domain pattern emblematic of the supercrystal phase. This is evidenced by the absence of the satellite peaks at the diagonal positions (white arrows, Figure 3b), features that were observed in measurements of the three-dimensional supercrystal phase[23]. The lack of checkerboard-domain patterning is consistent with the observation that the SS-x and -y phases occupy different real-space regions (Figure 2c).

Figure 3c shows the diffraction intensity integrated in the boxes along the respective Q-axes drawn in Figure 3b. The satellite peaks along the $Q_y$-axis are visible in the lineouts of VL-xy (green arrow) and SS-y phases (blue arrow). The first-order satellite peak of the VL-xy state (green arrow) subtly differs from the satellite peak of the vortex state (red arrow), meaning that there is also a subtle difference between their in-plane periodicities. This is consistent with the $Q_x$ lineouts showing that the second-order SS-x peaks (near the first-order pristine and VL-xy peaks) are also slightly shifted and correspond to a real space periodicity of ≈12 nm.

## 4. Phase-field Simulation and Outlook

To gain a deeper understanding of the formation and evolution of these polar phases, phase-field simulations were performed to emulate the low-fluence, multi-shot optical excitation (Methods). By tuning the optical dosage and the sample temperature, the VL-xy and SS phases were stabilized in the simulations (Figure 4a), consistent with the experiments. In the simulation of the *intermediate-dose* regime, it was found that the system transformed into a mixture of orthogonally arranged vortex-like tubes at a carrier dosage of $10^{27}$ m$^{-3}$ and a peak temperature of 550 K, in agreement with the observed VL-xy phase (Figure 4b). The microscopic polarization configuration of VL-xy phases features curled polarization to form vortex-like tubes. These tubes extend in both *x* and *y* axes and are truncated into shorter length than vortex tubes. The peak temperature here refers to the peak lattice temperature that follows the optical excitation, which subsequently dissipates via heat transport to the substrate. Notably, this charge density is comparable to what was previously used in phase-field simulations yielding the three-dimensional supercrystal phase[23] and is consistent with the transient processes of the supercrystal formation that were measured recently[41]. The key difference that lead to the VL-xy phase was to tune the peak-lattice temperature below 550 K, rather than 600 K which led to the supercrystal phase in previous work[23]. To simulate the experimental excitation conditions with accumulated exposure time, we further increased the charge concentration without changing the peak temperature. In this *high-dose* regime, neither supercrystal nor VL-xy phases were observed. Instead, the SS phases emerged featuring a wider stripe without any checkboard patterns. Although the simulation size is limited, the SS phase stripe period was found to be about double the periodicity of the VL-xy phase, as observed in our experiments (Figure 4b). Since the SS period is doubled, the polarization has room to accommodate orthogonal polarization without curling into vortex tubes. Therefore, the SS phases feature domains with polarization up and down, distinct from the in-plane polarizations in $a_1/a_2$ phase.

These simulations show that charge concentration and peak-lattice temperature following optical excitation are the key factors that govern the formation of the optically induced phases. Although the majority of photocarriers are recombined, the carriers can be trapped indefinitely to screen polarization locally, as indicated by the long structural dynamics in photoexcited ferroelectrics[42,43]. The simulation shows that the excited carriers that do not recombine after each pulse excitation migrate to domain boundaries, the PTO/STO interfaces, and regions with high electrostatic-energy density, spatially separating them and leaving these regions in charged states[16]. The process is compounded by the next optical excitation, adding more charge carriers to these local regions. The accumulated increase in local charge concentration in concert with a modest lattice temperature can thus lead to subtle and gradual changes in domain ordering templated by the existing phases. This repetitive excitation using multiple laser shots opens the opportunity to explore metastable states that are difficult to access otherwise.

The observed evolution of polar phases can be thought of as a particular trajectory through a light-induced "phase diagram" (Figure 1b), in which the energy surface was qualitatively based on the free energy of each polar phase obtained by the phase-field calculations (Table 2, Supporting Information). This phase diagram suggests new strategies for controlling polar nanostructures with light. For example, to obtain the VL-xy phase, the peak temperature needed to be kept below 550 K while still maintaining a relatively high charge concentration around $10^{27}$ m$^{-3}$. To obtain the SS phase, which has a higher energy and is not stable at ambient conditions, longer optical exposures that further increase the charge concentration without increasing the peak-lattice temperature can chart an isothermal path from the VL-xy phase to the SS phase (blue arrows, Figure 1b). In contrast, a single-shot high-fluence excitation that simultaneously raises the lattice temperature and carrier concentration above a certain threshold can lead to a different path (red arrow, Figure 1b) for creating the supercrystal phase.

## 5. Conclusion

Our work demonstrated the perspective of using PTO/STO SLs for simulating responses of neural networks by imaging the evolution of its domain networks upon cumulative ultrafast optical excitations. *In-situ* x-ray nanodiffraction quantitatively revealed the localized creation, transformation, and reconfiguration of light-induced metastable phases, providing the mechanistic insights into their adaptive responses to optical stimuli. The strategy of using a low-fluence, multi-shot optical excitation can also be applied to a wide range of solid-state systems for engineering nano- and micro-scale structures, opening a rich parameter space to explore and manipulate nanoscale domain networks. By programing the interval, intensity, and wavelength of optical pulse trains, the responses can be tuned to create, for example, light-programmable nanoscale electronic circuits or synaptic weights in a light-based neuromorphic device[44].

## 6. Experimental Section

*Sample Preparation*

(PbTiO$_3$)$_{16}$/(SrTiO$_3$)$_{16}$ superlattices, with $n$ = 16 unit cells were synthesized on a DyScO$_3$ (110) substrate via reflection high-energy electron diffraction (RHEED)-assisted pulsed laser deposition, using established procedures[9]. The thickness of the superlattice is 100 nm. These samples were composed of mixed vortex and $a_1/a_2$ phases with a mesoscale order of 400 nm, which is confirmed by x-ray diffraction (XRD) (see Figure 1) before exposing the sample to laser pulses.

*In-Situ X-Ray Nanodiffraction Imaging*

The experiment was performed at the Nanoprobe beamline at the Center for Nanoscale Materials of Argonne National Laboratory, using a similar setup reported previously[31]. A 10 keV focused x-ray beam was used to measure the 002 Bragg peaks of the vortex phase at the $m$ = -1 superlattice ordering peak. A Fresnel zone plate with a diameter of 160 µm and an outer zone width of 30 nm was used to focus the x-rays to a spot of 35 nm in diameter (full-width-half-maximum). Although the focused x-ray spot size is only 35 nm, the x-ray simultaneously probes at least two phases due to the shallow incident angle (Figure S2, Supporting Information). Due to the large divergence of the tightly focused x-ray beam (≈0.2 degrees), the large acceptance of the area detector, and the fact that reciprocal space has closely spaced peaks that cut through the detector in this diffraction geometry, a multitude of peaks in the shape of the annulus can be observed at this fixed incident angle. Instead of sharp peaks on the detector, Bragg peaks are shown as annuli because the pattern is a convolution of the shape of the zone plate with the Bragg diffraction peaks. The optical excitation is derived from a fiber laser (Model: Satsuma, Amplitude Inc.) running at a repetition rate of 67 kHz with a fundamental wavelength of 1030 nm and a pulse length of 400 fs. The fundamental light was tripled to produce 343 nm wavelength. The optical beam was focused to 200 µm in diameter (FWHM) via a pair of concave and convex lenses. The optical excitation is homogeneous because the pump beam size is an order of magnitude smaller than the probed region. With an incident angle of 20 degrees with respect to the sample surface, a maximum incident fluence of 2.3 mJ/cm$^2$ can be reached. The absorbed photon density can be calculated by considering optical reflection and absorption using the following equations: $n = (1 - R)(1 - e^{-\alpha L})F/(E_{h\nu}L)$, where the fluence $F$ is 2.3 mJ cm$^{-2}$, the photon energy of the 343 nm light $E_{h\nu}$ is 5.76×10$^{-19}$ J, the optical reflectivity $R$ is 0.6, PTO absorption coefficient $\alpha$ is 4×10$^5$ cm$^{-1}$ at 343 nm, PTO layer thickness $L$ is 50 nm. In this calculation, the light absorption in STO is negligible because its absorption coefficient 2×10$^4$ cm$^{-1}$ is 20 times smaller than that of PTO. Plugging the above numbers, the absorbed photon density per pulse is 2.8×10$^{26}$ m$^{-3}$. Compared to the threshold fluence of 30 mJ cm$^{-2}$ at 400 nm optical excitation[23], this is about 12% of the absorbed photon density required for creating three-dimensional supercrystal phases.

*Data Processing for Scanning X-Ray Diffraction Images*

The real-space imaging data sets were first corrected for translation drifts by cross-correlating the total integrated intensity map of each exposure. Only the area where all the maps overlap was used, reducing the data set from 80000 to 30976 diffraction images. The total integrated intensity maps were further corrected to the sub-pixel level by finding the affine transformation that minimized the mean-squared error (MSE) via an iterative automatic differential optimizer. The diffraction patterns of each exposure time were summed (Figure S6b, Supporting Information). Based on the total integrated intensity maps and the summed diffraction patterns, three optical regimes were deduced. They are the following: pristine – 1 s, when no new nanoscale domain arrangements appeared, 5 – 10 s, when the VL-xy phases appeared (based on the "cross-hatching" domain ordering appearing), and 50 – 1000 s, when the superstripe phases appeared (based on the new features appearing in the summed diffraction patterns). Diffraction patterns were then convolved with a 5 × 5 uniform kernel in the plane of the detector to increase the signal to noise.

For the low-dose regime (pristine-1s), there are only 2 crystallographic species present in the pristine and 1 s data sets: the vortex and $a_1/a_2$. The satellite intensity is a reliable signature to differentiate the vortex phases from the $a_1/a_2$ phase. We thus use the diffraction intensity of the in-plane satellite in a region of interest (ROI), which was then plotted as a function of real space. A threshold of 44% of the normalized intensity was used to determine if a diffraction pattern in the pristine or 1 s exposure data set was predominantly the vortex phase or the $a_1/a_2$ phase. For the pristine diffraction patterns in Figure 2b, the 5 diffraction patterns with the brightest satellites and the dimmest satellites according to the in-plane satellite ROI were averaged to create the vortex and $a_1/a_2$ diffraction patterns respectively.

For the intermediate dose regime (5 – 10 s), k-means clustering with 3 categories was used[37,45]. One of the categories was the "VL-xy" state, which displayed diffraction patterns that were distinct from the pristine state. The regions with VL-xy-like diffraction were designated as such. The remaining real-space positions, composed of vortex and $a_1/a_2$ phases, were analyzed using the same satellite fitting threshold procedure described above to separate the regions of vortex and $a_1/a_2$ phases. The VL-xy diffraction shown in Figure 2b was created by taking 100 diffraction patterns labeled VL-xy in the 10 s exposure series and averaging them.

For the high-dose regime (50 -1000 s), the logarithm of the convolved diffraction patterns was taken to optimize the k-means output. Four categories were found to be appropriate to categorize the diffraction patterns associated with 50 seconds of exposure, and 3 categories for 100, 500, and 1000 seconds of exposure[37,45]. The categorization using k-mean clustering in this regime produced groups of diffraction patterns with scattering features that can be related to the distinct species (Figure S3, Supporting Information). The 100 brightest summed ROI intensity diffraction patterns of the blue ROI and the purple ROIs were averaged to produce the SS-y and SS-x diffraction patterns, respectively, in Figure 2b.

Figure 2c was created by assigning every real space position on the cross-correlated data sets an integer and color based on which of the 5 different phases each diffraction pattern belongs. The affine transformation matrices used for Figure 2b were then applied to the domain maps, using a nearest-neighbor interpolation. The boundary maps (Figure 2d) were created by taking the Euclidean distance of the vertical and horizontal gradients with respect to the classification number of the corrected domain maps. Anything > 0 was considered a boundary.

To assess the uncertainty of phase assignment, we compare a second method, which uses satellite intensity to assign phase, against K-means clustering. We focus on the SS-y phase because its distinct satellite peaks are well separated from other peaks (Figure S8, Supporting Information). The two methods disagree on 5.5% of pixels, providing a metric for SS-y phase assignment uncertainty. Although other phases are more difficult to evaluate due to more complex diffraction signatures, this example offers a quantitative uncertainty estimate.

*Three-Dimensional Reciprocal Map Construction*

A procedure used in previous works[46] was used to create the three-dimensional reciprocal space maps (RSMs) shown in Figure 3. Due to the experimental difficulty of illuminating a constant nanoscale volume during the sample rotation, we intentionally collected the diffraction patterns in Figure 3 as a function of the sample position (*x*) along the x-axis and the sample rotation angle ($\theta$) to average the polar phases of interests over micrometers (Figure S7, Supporting Information). For the pristine and intermediate phases, a $\theta$ range from 15.3 to 21.3 degrees with steps of 0.1 degrees and an *x* range of 4 µm with 200 nm steps were used. For the superstripe RSM, a $\theta$ range from 16.3 to 20.8 degrees with steps of 0.1 degrees and an *x* range of 10 µm with 500 nm steps were used. The RSMs, therefore, visualize the unique diffraction features and quantitatively measure the out-of-plane lattice constant of the pristine (i.e., vortex and $a_1/a_2$), VL-xy, and SS phases.

The detector images were then converted into reciprocal space coordinates ($Q_x$, $Q_y$, $Q_z$) using equations and procedures in previous works for an out-of-plane Bragg peak, which depends only on the experimental geometrical parameters[46]. Here, the pixel size was 75 µm, the distance of the detector from the sample was 0.646 m, and the x-ray wave vector magnitude was $2\pi/\lambda = 5.06$ Å$^{-1}$. The intensity as a function of reciprocal space coordinates was then interpolated to make a uniform, three-dimensional grid, where each three-dimensional position contains an intensity.

*Phase-Field Simulation*

We model the microstructure and light-induced transition of the PbTiO$_3$/SrTiO$_3$ superlattice system by solving the temporal evolution and spatial distribution of a set of independent variables, including the ferroelectric polarization **P**, the mechanical displacement **u**, the electron and hole concentrations $n$ and $p$, and the electric potential $\Phi$. The time-dependent evolution of these variables is described by a set of equations, including the polarization dynamics equation, the elastodynamics equation, the electron and hole transport equations, and the electrostatic Poisson equation, written as[47,48]:

$$\boldsymbol{\mu}\ddot{\mathbf{P}} + \boldsymbol{\gamma}\dot{\mathbf{P}} = -\delta F/\delta \mathbf{P}, \tag{1}$$
$$\rho\ddot{\mathbf{u}} = \nabla \cdot (\boldsymbol{\sigma} + \beta\dot{\boldsymbol{\sigma}}), \tag{2}$$
$$\dot{n} = -\nabla \cdot \mathbf{j}_e + R, \tag{3}$$
$$\dot{p} = -\nabla \cdot \mathbf{j}_h + R, \tag{4}$$
$$\nabla \cdot (-\kappa_0 \boldsymbol{\kappa}^b \nabla \Phi + \mathbf{P}) = e(p - n). \tag{5}$$

An overdot indicates the derivative over the time, and a double overdot indicates the second-order time derivative. In Equation 1, **µ** and **γ** are the mass and damping coefficients of the polarization and $F$ is the total free energy of the system. In Equation 2, $\rho$ and $\beta$ are the material mass density and the stiffness damping coefficient, respectively, and **σ** is the stress field. In Equations 3-4, $\mathbf{j}_e$ and $\mathbf{j}_h$ are the net fluxes of free electrons and holes, respectively and $R(\mathbf{x})$ is the generation rate of electron-hole pairs which includes light-induced excitation and spontaneous electron-hole creation and recombination. In Equation 5, $\kappa_0$ and $\boldsymbol{\kappa}^b$ are the vacuum permittivity and background dielectric constant, respectively, and $e$ is the elementary charge. We refer to Ref. [48] for more details of the phase-field model including the free energy functional, the stress field expressions, and the net fluxes and generation rates of electrons and holes. By numerically solving equations (1-5), we obtain the spatiotemporal response of the coupled electron-lattice system during light-induced transitions.

The simulation system contains 2 PbTiO$_3$ layers and 2 SrTiO$_3$ layers, where each layer has a thickness of 4.8 nm, equivalent to 12 unit-cell lengths of the perovskite crystal lattice. The in-plane dimensions of the system are chosen as $80 \text{ nm} \times 80 \text{ nm}$. 3-dimensional periodic boundary conditions are employed for polarization, strain, carrier concentrations, and electric potential. The spatial averages of the out-of-plane stress components $\sigma_{13}$, $\sigma_{23}$, and $\sigma_{33}$ are fixed at zero for describing an out-of-plane stress-free condition, while the spatial averages of the in-plane strain components are fixed at[16] $\varepsilon_{11} = -0.003$, $\varepsilon_{22} = 0$, and $\varepsilon_{12} = 0$. The material constants used in the phase-field simulations are the same as those in Ref. [16], which are taken from Ref.[48–52].

The pristine state of the PbTiO$_3$/SrTiO$_3$ superlattice with a mixture of $a_1/a_2$ and vortex domains is obtained by evolving the system to an equilibrium state at room temperature. Using the pristine state as an initial condition, we simulate the light-induced formation of the VL-xy structure as well as the final SS structures by considering both the heating and carrier excitation effects of the light irradiation. For simulating the formation of the VL-xy phase in the low-dose regime, we employ a light-induced carrier generation rate that instantly raises the concentrations of both free electrons and holes to $10^{27}$ m$^{-3}$, which is comparable to the estimated values for inducing supercrystals in our previous study[23]. Simultaneously, we consider a rise of the temperature to 550 K and a subsequent relaxation back to room temperature as a result of heat transport to the substrate, representing a process of lattice heating and cooling following optical excitation. Both the light-induced charge generation rate and the temperature change are considered homogeneous for describing spatially uniform light irradiation. The formation of the VL-xy phase is then simulated by evolving the electron-lattice system under these laser-induced effects. We note that the charge concentration used in the simulation is three orders of magnitude smaller than the absorbed photon density in the experiments, suggesting that most of the excited electron-hole pairs recombine before the next pulse arrives and only about 0.1% of excited carriers were trapped.

In the high-dose regime, a continuous carrier injection at a constant rate was employed to simulate the high-repetition laser shots. The formation of the SS phases was observed by applying a total dosage of $10^{30}$ m$^{-3}$ electron-hole pairs. The resulting calculated x-ray diffraction pattern[53] agrees qualitatively with the experimental results. Experimentally, this high dose condition is achieved by spreading the total dosage into millions of pulses at a high repetition rate of 67 kHz so that the carriers do not fully recombine and can accumulate. The simulations further indicate a spatially averaged carrier concentration of $10^{24}$ m$^{-3}$ when the SS phase is stabilized. The effect of the steady-state lattice temperature the system cools down to after each laser shot was further examined. It was found that the SS structure forms under a temperature increase smaller than 20 K, while the structure is destroyed under larger temperature changes.

**Acknowledgments**
We acknowledge the discussion with Aaron M. Lindenberg. This work was supported by the U.S. Department of Energy, Office of Science, Basic Energy Sciences, Materials Sciences and Engineering Division (experimental design, data collection and analysis, manuscript preparation by M.Z., Y.C., S.H., H.W. were supported under Contract No. DE-AC02-06CH11357; sample preparation and part of data interpretation by V.S., L.M., V.G., J.F., H.W were supported under contract number DE-SC0012375). The phase-field simulations were supported as part of the Computational Materials Sciences Program funded by the U.S. Department of Energy, Office of Science, Basic Energy Sciences, under Award No. DE-SC-0020145. Work performed at the Center for Nanoscale Materials and the Advanced Photon Source (B.G., M.C, M.H.), both U.S. Department of Energy Office (DOE) of Science User Facilities, was supported by the U.S. DOE, Office of Basic Energy Sciences, under Contract No. DE-AC02-06CH11357. S.D. and R.R. acknowledge support for part of sample synthesis through the Quantum Materials program (KC 2202) funded by the U.S. Department of Energy, Office of Science, Basic Energy Sciences, Materials Sciences Division under contract no. DE-AC02-05-CH11231.

**Supporting Information**
Supporting Information is available from the Wiley Online Library or from the author.

**Conflict of Interest:**
The authors declare no conflict of interest.

**Data Availability Statement:**
The data that support the findings of this study are available from the corresponding author upon reasonable request.

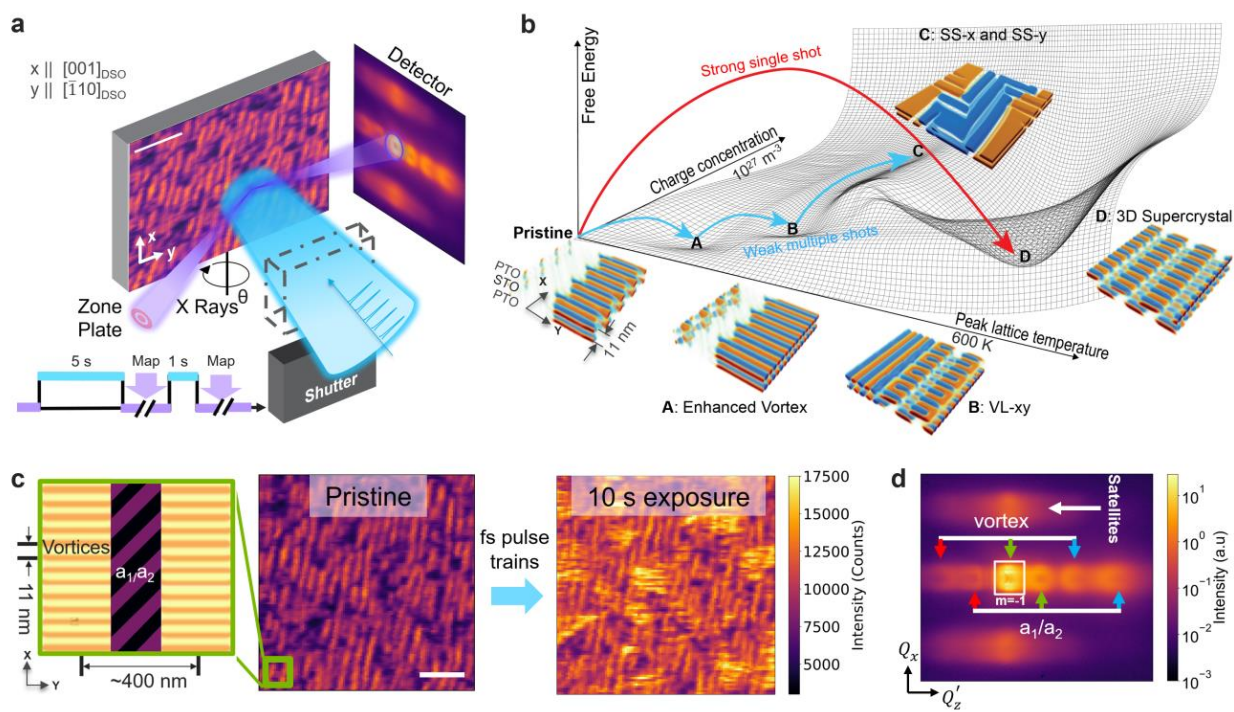

**Figure 1 Optical control of nanoscale polar structures.** a) Synchrotron-based *in-situ* nanodiffraction imaging with laser excitation. The inset at the lower left-hand corner shows the time sequence of x-ray imaging and light exposure. The laser shutter remained closed while the sample was mapped by focused x-rays, and then was opened for light exposure for the desired time. b) A qualitative free energy surface based on phase-field calculations (Supporting Information, Table 2). It illustrates that tailored optical excitations can explore the parameter space spanned by peak lattice temperature and accumulated charge density for accessing the metastable states. The insets show the nanoscale ordering of a trilayer (PTO/STO/PTO) model from the phase-field simulation. The orange and blue colors represent regions in which the polarization is aligned up or down, respectively. The axes labels are not in scale. c) The real-space maps of the diffraction intensity in the region of interest (ROI, white box in (d)) in the pristine and after 10-second optical exposure. The white scale bars represent 2 µm. The zoom-in left inset illustrates the schematic nanoscale domain configuration (not data). d) Summed diffraction pattern showing the vortex and the $a_1/a_2$ superlattice peaks. The central horizontal row of annuli along the $Q_z'$ direction on the detector, originating from regions containing vortex domain ordering and from $a_1/a_2$ domain arrangements. The green arrows indicate the $m = -1$ order SL peaks of the vortex-containing and the $a_1/a_2$ – containing regions. Diffuse scattering originating from the ordering of the vortex phase is also present, displaced from the CTR by $Q_x = 0.059$ Å$^{-1}$ vertically.

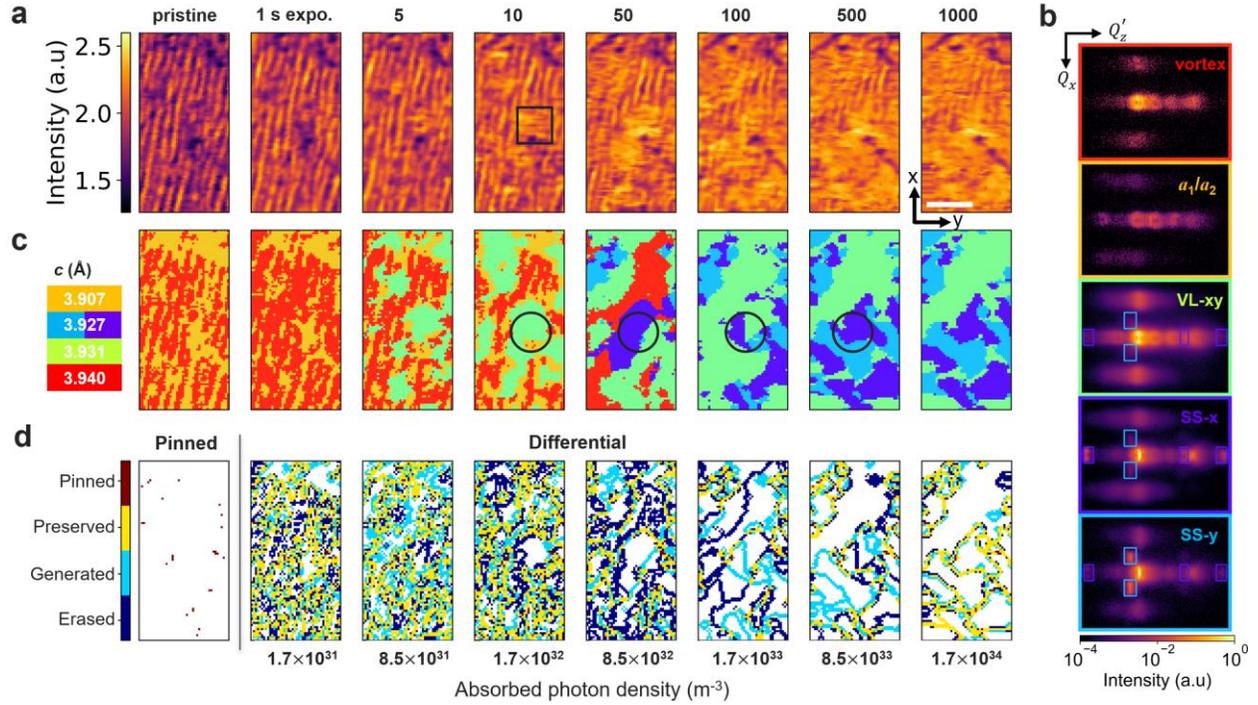

**Figure 2 Real-space imaging of phase evolution as a function of optical exposure time.** a) Maps of the -1 SL intensity after exposed to a pulse train with designated time (top label) and an incident fluence of 2.3 mJ cm$^{-2}$. The corresponding absorbed photon density is labelled at the bottom of (d). The black box highlights a representative region of horizontal streaks. b) Diffraction patterns of five categorized phases taken at $\theta$=18.3°. c) Maps of phase distribution as a function of optical exposure time. They are color-coded corresponding to their diffraction patterns shown in (b). The white scale bar is 2 µm. The black circles highlight regions where the phases interconvert. d) Maps of domain boundaries. The pinned map shows the distribution of domain boundaries that survived the whole optical exposure series in (a). The differential maps illustrate the change of the phase boundaries. For example, the differential map under 1s exposure is derived from the difference between 1s and pristine maps in Figure S5, Supporting Information. Yellow, light blue, and dark blue color denote a boundary that was respectively, persisted, created, or erased upon the subsequent light exposures in series.

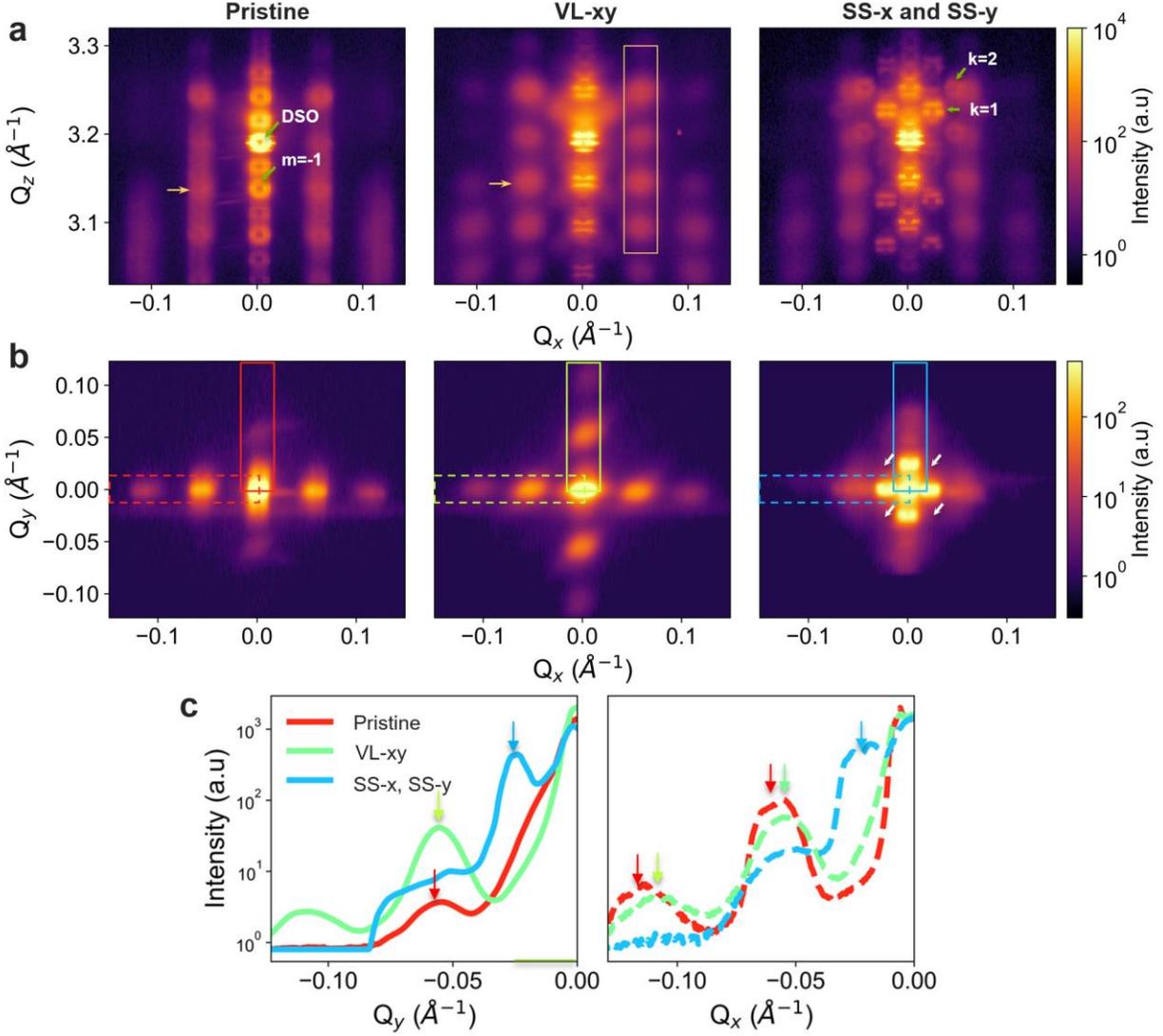

**Figure 3**. **Reciprocal maps that primarily contain pristine, VL-xy, and SS phases.** a) $Q_z – Q_x$ reciprocal map over a slice in $Q_y \in$ (-0.02, 0.02) Å$^{-1}$. The yellow arrows indicate the m=-1 satellite peaks of the vortex and VL-xy phase. The green arrows indicate the substrate (DSO) peak, m=-1 SL peak of vortex phase, the first (k=1) and second (k=2) order satellite peaks of the SS-x phase. The vertical yellow box in the VL-xy panel indicates the ordered satellite peaks comparing with the pristine phase. b) $Q_x – Q_y$ reciprocal map integrated a slice in $Q_z \in$ (3.125, 3.15) Å$^{-1}$ for pristine, (3.125, 3.165) Å$^{-1}$ for VL-xy, and (3.21,3.24) Å$^{-1}$ for SS phases, respectively. The white arrows indicate the positions of missing diffraction peaks that are present in the supercrystal phase. c) Lineouts of the reciprocal space maps corresponding to the colored boxes averaged along the corresponding dimensions shown in (b).

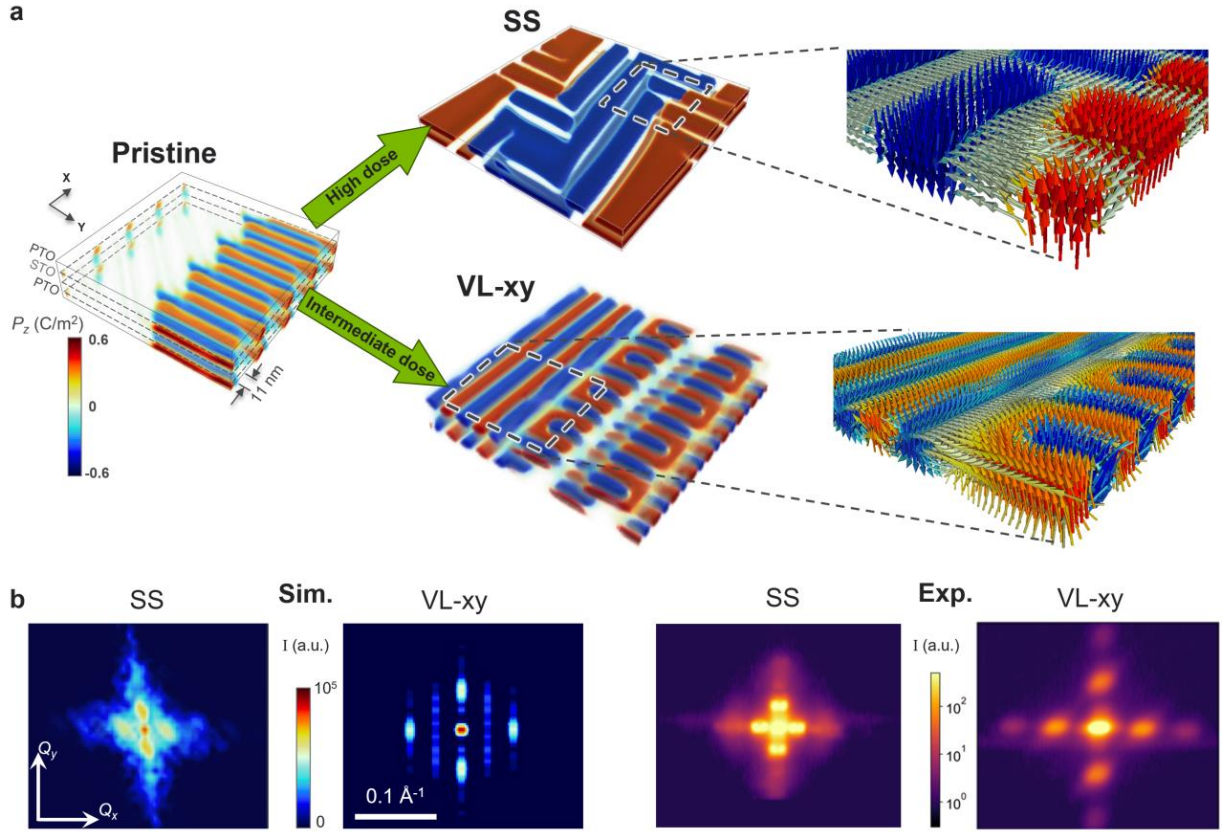

**Figure 4. Phase-field simulations.** (a) VL-xy and SS phases emerge from the pristine phase under intermediate and high optical dose conditions, respectively. The zoomed-in graphs show the polarization configurations of VL-xy and SS phases. The white arrows indicate the polarization lying in the x-y plane, similar to the polarization in the $a_1/a_2$ phase. The red and blue arrows show the up and down polarization, respectively. (b) XRD simulation of the real-space domain patterns in (a). Tilting of the simulated SS pattern away from the $Q_y$ axis is due to the limited simulation size.

# Supporting Information for

## Optical Control of Adaptive Nanoscale Domain Networks

Marc Zajac[1], Tao Zhou[2], Tiannan Yang[3,11], Sujit Das[4,12], Yue Cao[5], Burak Guzelturk[1], Vladimir Stoica[3], Mathew Cherukara[1], John W. Freeland[1], Venkatraman Gopalan[3], Ramamoorthy Ramesh[4,7,8,9], Lane W. Martin[4,7,8,9,10], Long-Qing Chen[3], Martin Holt[2], Stephan Hruszkewycz[5*], Haidan Wen[1,5*]

[1] Advanced Photon Source, Argonne National Laboratory, Lemont, IL 60439
[2] Center for Nanoscale Materials, Argonne National Laboratory, Lemont, IL 60439
[3] Department of Materials Science and Engineering, Pennsylvania State University, University Park, PA 16802
[4] Materials Sciences Division, Lawrence Berkeley National Laboratory, Berkeley, CA 94720
[5] Materials Science Division, Argonne National Laboratory, Lemont, IL 60439
[7] Department of Materials Science and NanoEngineering, Rice University, Houston, TX 77005
[8] Department of Physics and Astronomy, Rice University, Houston, TX 77005
[9] Rice Advanced Materials Institute, Rice University, Houston, TX 77005
[10] Department of Chemistry, Rice University, Houston, TX 77005
[11] Present address: Interdisciplinary Research Center, School of Mechanical Engineering, Shanghai Jiao Tong University, Shanghai, 200240, China
[12] Present address: Materials Research Centre, Indian Institute of Science, Bangalore, 560012, Karnataka, India

*Email: shrus@anl.gov; wen@anl.gov

**Supporting Figures**

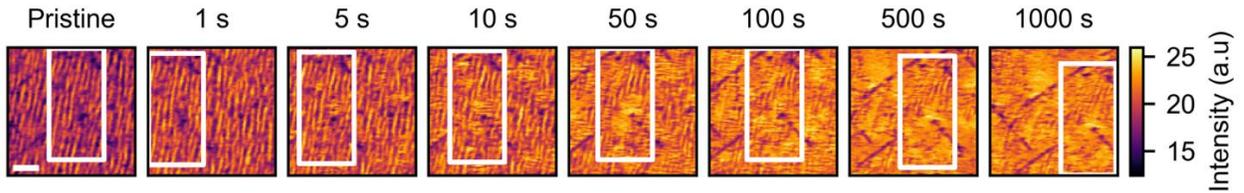

**Figure S1**: 10 µm x 10 µm intensity maps taken at the different exposure times. Features such as defect sites have been used as fiducial marks for drift correction. The white rectangles are drift-corrected regions that are compared and analyzed in Figure 2. The drift is mainly due to the thermal instability of the sample position with respect to the x-ray probe beam. The white scale bar in the lower-left corner of the pristine map represents 2 µm.

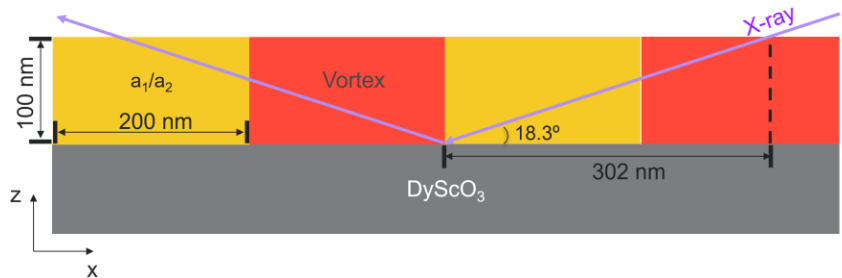

**Figure S2**: Schematic of the experimental geometry. The red and yellow represent the vortex and $a_1/a_2$ domains in the sample. The probed region by the x-ray (purple line) measures a phase mixture. The fraction of vortex domains probed in the pristine phase varies between 33%-67% as the x-ray beam is raster scanned across the sample. The real-space resolution of the x-ray probe is about 300 nm along the $x$ axis, while the resolution along the $y$ axis is limited to the focus size of 35 nm.

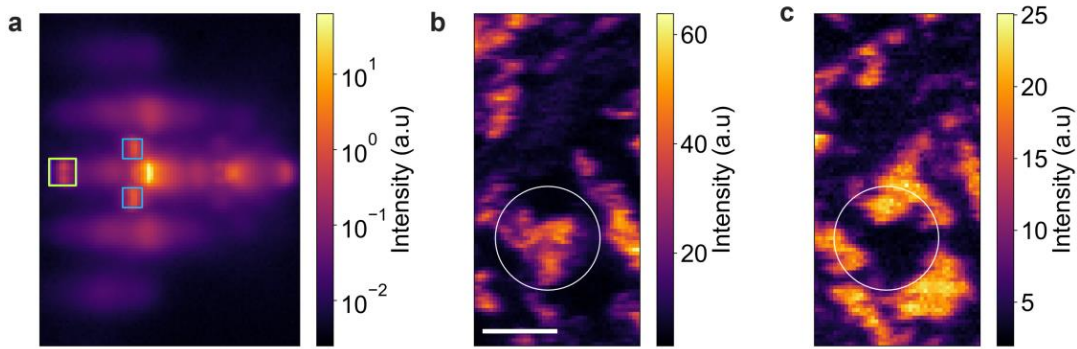

**Figure S3**: (a) Averaged diffraction pattern over the map taken after 1000 s optical exposure. (b) Integrated intensity map of SS-y diffraction in the light blue regions of interest (ROI) in (a). (c) Integrated intensity map of SS-x diffraction in the green ROI in (a). The white circle highlights a real-space region that is bright in (b) but dark in (c), suggesting that superstripe x and y phases occupy distinct regions in real space. The white scale bar is 2 µm.

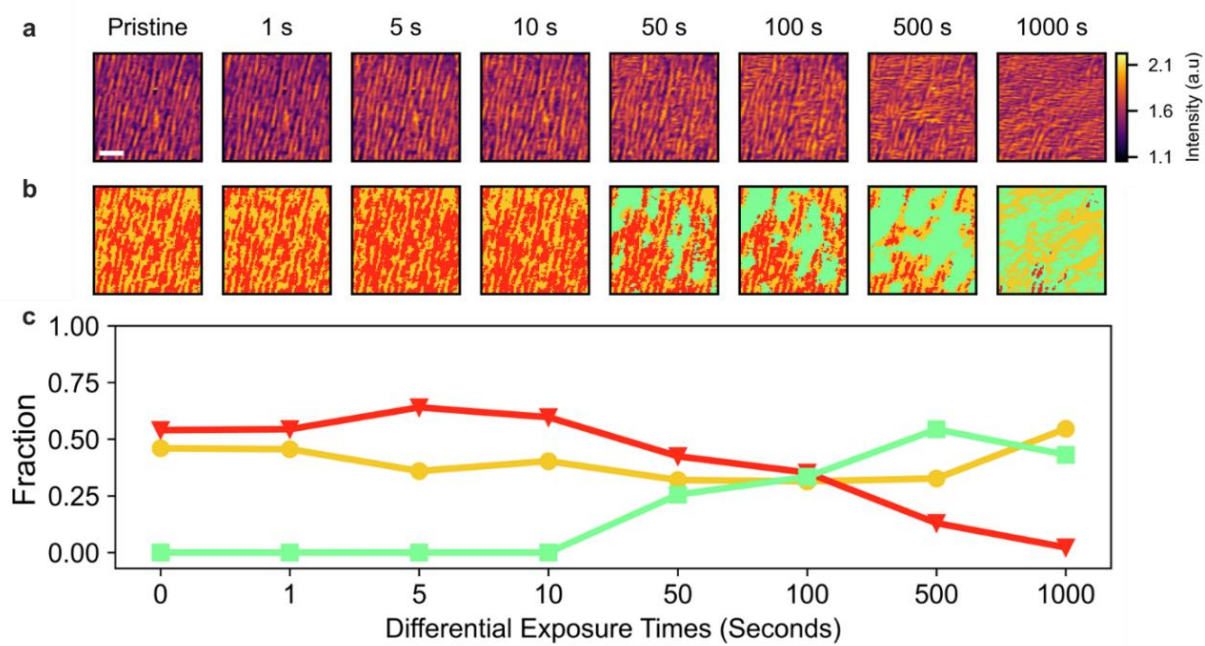

**Figure S4**: Exposure series with an incident fluence of 1.2 mJ cm$^{-2}$. a) Real-space maps of the -1 SL intensity. b) At half of the optical fluence used compared with the experimental condition in Figure 2, only three phases were observed based on the categorization using the same color code as in Figure 2c. The VL-xy phase started to appear after 50s of exposure rather than 10s, indicating a higher accumulated optical dose is needed to create the VL-xy phase. c) Fractional population of three phases as a function of optical exposure time. The scale bar in the pristine real space map is 2 µm. The final exposure times do not show any superstripe species, suggesting that the critical cumulative bound charge carrier density was not reached. No super-stripe regions were observed even after 1000 s exposure, while the superstripe phases appeared just after 50s exposure with a fluence of 2.3 mJ cm$^{-2}$ (Figure 2c). Thus, the threshold for creating the super-stripe phase does not scale linearly with the optical dose, indicating that the conversion is a nonlinear process.

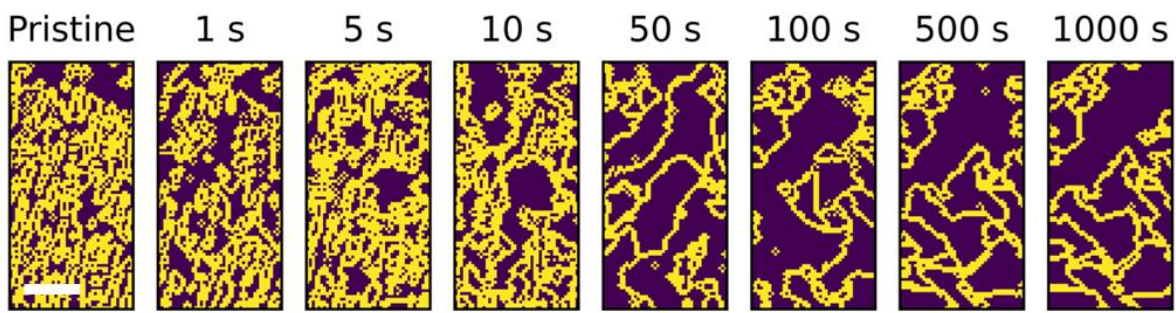

**Figure S5**: Maps of domain boundary (yellow) as a function of optical exposure time at a fluence of 2.3 mJ cm$^{-2}$. The scale bar in the pristine real space map is 2 µm.

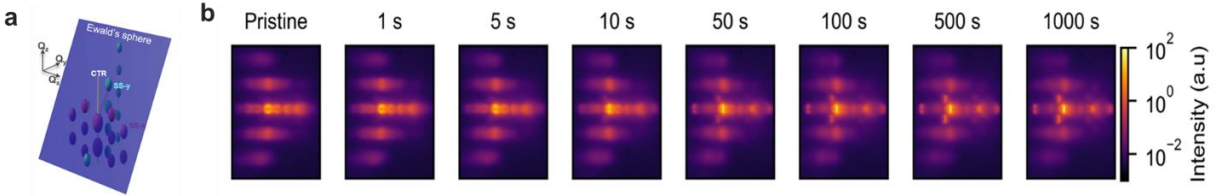

**Figure S6**: (a) Schematic of Ewald's sphere (blue plane) cutting through the reciprocal space of the superstripe phase. For clarity, only the diffraction spots intersect or below Ewald's sphere are shown. (b) The diffraction patterns summed over all scanning probed locations from the real-space regions in Figure 2.

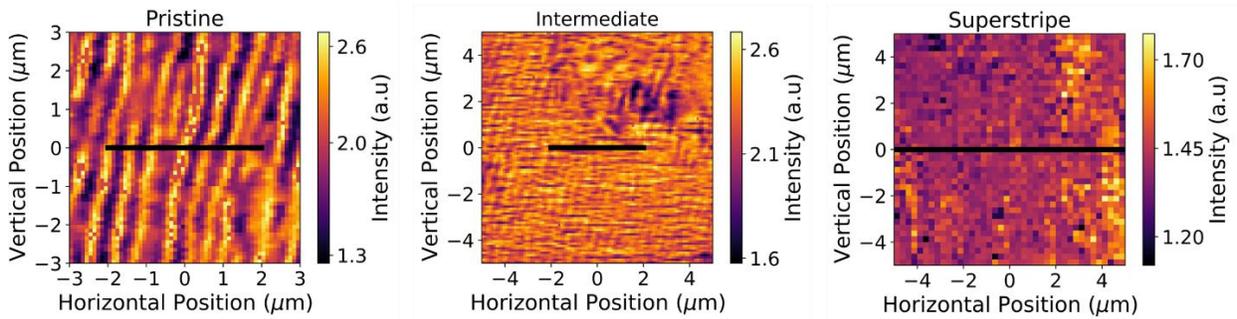

**Figure S7**: Integrated intensity maps of $m = -1$ Bragg peak where the reciprocal space maps in Figure 3 were taken. The one on the left was taken on a pristine part of the sample. The one in the middle was taken after being exposed at an average fluence of 1.2 mJ cm$^{-2}$ for 720 s. The one on the right was taken after being exposed at 1000 s with an average fluence from 2.3 mJ cm$^{-2}$. The black lines indicate the measured regions during the rocking ($\theta$) and translation (x) 2D scans.

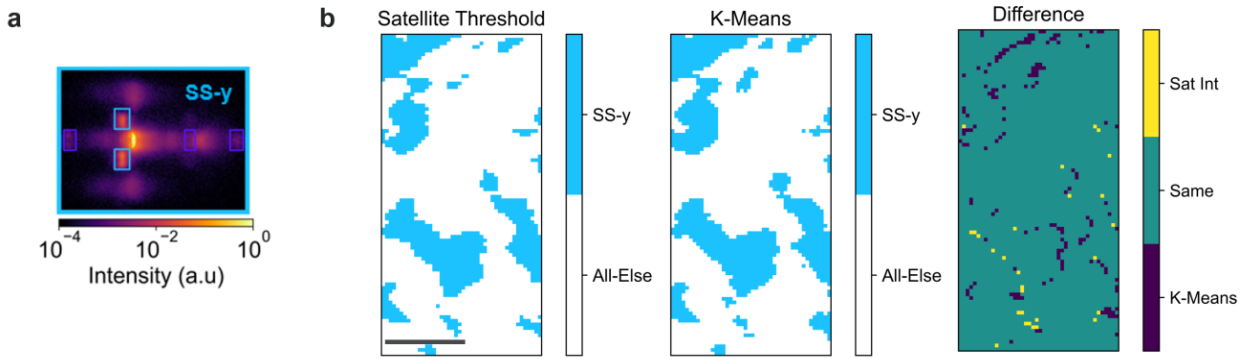

**Figure S8**: (a) Diffraction pattern that shows the satellite peak in light blue box (ROI) as a signature of SS-y phase. We first normalized the ROI intensity for all diffraction patterns. If the ROI intensity larger than 30% of the normalized intensity for a given pixel, it is categorized as a SS-y phase, otherwise, it is labeled "All-Else". (b) Maps of the SS-y phase for 1000 s exposure image in Fig. 2c determined using the satellite threshold method is compared with the K-Means

clustering. The difference map shows the potential miscategorized pixels, which is 4.7% of the total number of pixels. A similar procedure was also applied to 50, 100 and 500 s data sets. We found that the average percentage of number of pixels that the two methods disagree is 5.5%, which is used here to evaluate the uncertainty for assigning SS-y phase using K-means clustering. The black scale bar in the lower left-hand corner of the satellite threshold method plot represents 2 µm.

**Supporting Tables:**

| Pseudocubic (Å) | vortex | $a_1/a_2$ | VL-xy | SS-x | SS-y | $DyScO_3$ |
|---|---|---|---|---|---|---|
| $a$ | 3.952 [1] | 3.952 [1] | N/A | N/A | N/A | 3.952[4] |
| $b$ | 3.935 [1] | 3.952 [1] | N/A | N/A | N/A | 3.947[4] |
| $c$ | 3.94 [2,3] | 3.907 [2], 3.915[3] | 3.931[2] | 3.927[2] | 3.927[2] | 3.947[4] |

**Table 1**: Lattice constants of various polar structures.

[1] Li, Q. *et al.* Subterahertz collective dynamics of polar vortices. *Nature* **592**, 376–380 (2021).
[2] Measured by x-ray diffraction in this work.
[3] Damodaran, A. R. *et al.* Phase coexistence and electric-field control of toroidal order in oxide superlattices. *Nat. Mater.* **16**, 1003–1009 (2017).
[4] Schmidbauer, M. et al. High-precision absolute lattice parameter determination of $SrTiO_3$, $DyScO_3$ and $NdGaO_3$ single crystals, *Acta. Crystallogr. B.* **68**, 8 (2012).

| Polar Phases | Pristine | Enhanced Vortex | VL-xy | SS | Supercrystal |
|---|---|---|---|---|---|
| Relative Free energy ($\times 10^6$ J m$^{-3}$) | 0 | 1.01 | 1.55 | 2.83 | 1.65 |

**Table 2**: Relative free energies of the polar phases calculated from the phase-field simulations.